\documentclass[preprint,12pt]{elsarticle}

\usepackage{amssymb}
\usepackage{hyperref}
\usepackage{xcolor}
\usepackage[ruled,linesnumbered]{algorithm2e}
\journal{Chinese Physics C}

\begin{document}


\title{A Versatile Framework for Analyzing Galaxy Image Data by Implanting Human-in-the-loop on a Large Vision Model}

\author[1,3,4]{Ming-Xiang Fu\fnref{label1}}{}
\author[2]{Yu Song\fnref{label1}}{}
\author[2]{Jia-Meng Lv\fnref{label1}}{}
\author[2]{Liang Cao}{}
\author[2]{Peng Jia\corref{cor1}}{}
\author[1,4,3]{Nan Li\corref{cor2}}{}
\author[5]{Xiang-Ru Li}{}
\author[1,4]{Ji-Feng Liu}{}
\author[6,3,7]{A-Li Luo}{}
\author[8]{Bo Qiu}{}
\author[9]{Shi-Yin Shen}{}
\author[15]{Liang-Ping Tu}{}
\author[10]{Li-Li Wang}{}
\author[11]{Shou-Lin Wei}{}
\author[12]{Hai-Feng Yang}{}
\author[13]{Zhen-Ping Yi}{}
\author[14,7]{Zhi-Qiang Zou}{}
\cortext[cor1]{robinmartin20@gmail.com}
\cortext[cor2]{nan.li@nao.cas.cn}
\fntext[label1]{These authors contributed to the paper equally.}

\address[1]{National Astronomical Observatories, Chinese Academy of Sciences,Beijing 100101, China}

\address[2]{College of Electronic Information and Optical Engineering, Taiyuan University of Technology,Taiyuan 030024, China}

\address[3]{School of Astronomy and Space Science, University of Chinese Academy of Sciences,Beijing 101408, China}

\address[4]{Key lab of Space Astronomy and Technology, National Astronomical Observatories,Beijing 100101, China}

\address[5]{School of Computer Science, South China Normal University,Guangzhou 510631, China}

\address[6]{CAS Key Laboratory of Optical Astronomy, National Astronomical Observatories,Beijing 100101, China}

\address[7]{University of Chinese Academy of Sciences, Nanjing,Nanjing 211135, China}

\address[8]{University of Science and Technology Beijing,Beijing 100083, China}

\address[9]{Shanghai Astronomical Observatory, Chinese Academy of Sciences,Shanghai 200030, China}

\address[10]{School of Computer and Information, Dezhou University,Dezhou 253023, China}

\address[11]{Faculty of Information Engineering and Automation, Kunming University of Science and Technology,Kunming 650500, China}

\address[12]{School of Computer Science and Technology, Taiyuan University of Science and Technology,Taiyuan 30024, China}

\address[13]{School of Mechanical, Electrical and Information Engineering, Shandong University,Weihai 264209, China}

\address[14]{Nanjing University of Posts \& Telecommunications,Nanjing 210023, China}

\address[15]{School of Science, university of Science and Technology LiaoNing,Anshan 114051, China}


\begin{abstract}
The exponential growth of astronomical datasets provides an unprecedented opportunity for humans to gain insight into the Universe. However, effectively analyzing this vast amount of data poses a significant challenge. Astronomers are turning to deep learning techniques to address this, but the methods are limited by their specific training sets, leading to considerable duplicate workloads too.  Hence, as an example to present how to overcome the issue, we built a framework for general analysis of galaxy images, based on a large vision model (LVM) plus downstream tasks (DST), including galaxy morphological classification, image restoration, object detection, parameter extraction, and more. Considering the low signal-to-noise ratio of galaxy images and the imbalanced distribution of galaxy categories, we have incorporated a Human-in-the-loop (HITL) module into our large vision model, which leverages human knowledge to enhance the reliability and interpretability of processing galaxy images interactively. The proposed framework exhibits notable few-shot learning capabilities and versatile adaptability to all the abovementioned tasks on galaxy images in the DESI legacy imaging surveys. Expressly, for object detection, trained by 1000 data points, our DST upon the LVM achieves an accuracy of $96.7\%$, while ResNet50 plus Mask R-CNN gives an accuracy of $93.1\%$; for morphology classification, to obtain AUC $\sim0.9$, LVM plus DST and HITL only requests 1/50 training sets compared to ResNet18. Expectedly, multimodal data can be integrated similarly, which opens up possibilities for conducting joint analyses with datasets spanning diverse domains in the era of multi-message astronomy.
\end{abstract}

\maketitle



\section{Introduction}\label{section1}
The vast data expansion has positioned it as an invaluable resource across various scientific disciplines, notably physics and astronomy, which brings opportunities and challenges for human beings to understand the Universe. Artificial intelligence (AI) techniques have emerged as a leading approach to comprehending the complexities intrinsic to scientific data to overcome the big data challenge in physics, such as data collected from large-scale sky surveys, gravitational waves detectors, and colliders are more than an order of magnitude larger in size but even require a shorter processing time to respond to instant events promptly~\cite{2023arXiv230608106A}, and achieved significant success, for instance, predicting multivariate time series data drawn from particle accelerators~\cite{PhysRevAccelBeams.26.024801} , doing many-body variational calculations in nuclear physics~\cite{RevModPhys.94.031003}, and many other areas in experimental and theoretical physics (see ~\cite{2024FrP....1222162S} and references therein). \\

The big data challenge in astronomy and astrophysics is noteworthy especially \cite{zhang2015astronomy}, since large-scale sky surveys such as LSST\footnote{\url{https://www.lsst.org/}}, Euclid\footnote{\url{https://www.euclid-ec.org/}}, CSST\footnote{\url{http://nao.cas.cn/csst/}}, and SKA\footnote{\url{https://www.skao.int/}} continue to advance, leading astronomy and astrophysics to an exciting new era. However, the vast and intricate nature of astronomical data sets poses a significant challenge to astronomers who want to extract meaningful scientific information. Deep learning techniques have been used to address the problems (see~\cite{HCM2023} and references therein), for example, astronomers leverage specific data in supervised learning to teach computers how to solve problems, which has been successful in detecting celestial objects~\cite{lao2021artificial}, classifying their morphology~\cite{banerji2010galaxy,wu2019radio} and the spectrum~\cite{li2018carbon,xu2020two}. Also, unsupervised learning algorithms can explore unlabeled data and have shown effectiveness in classifying galaxy types~\cite{martin2020galaxy,logan2020unsupervised,stad3181,2021ApJ...911L..33H} and characterizing (or improving) the performance of telescopes~\cite{jia2017blind,wang2018automated,2022ApJ...934...83N,2023MNRAS.525.5278G}. In addition, reinforcement learning algorithms have succeeded in various applications, such as efficiently managing instruments by developing simulators and enabling interactions with observations~\cite{jia2023observation,jia2023simulation}.\\ 

However, for the machine learning based applications discussed above, some issues still need to be addressed, including interpretability, data labeling, and universality. Persistent issues that hinder their advancement and utility require preparing separate training sets and constructing distinct models for different tasks. Despite this, various tasks may share a common foundation of prior information about celestial objects. For example, tasks like detecting strong gravitational lensing systems, different types of nebulae or galaxies, or segmenting galaxies share the need for multi-color structural features. Therefore, creating a foundation model that provides general information and attaches subprocesses for multiple different purposes is sensible. Besides, effectively training a machine learning algorithm typically requires thousands of data units, further problematizing matters, as specific data and labels are complex to obtain, such as the positions of rare astronomical targets or segmentation labels for galaxies. Therefore, an interactive manner is ideal for building training sets from scratch and keeping them developing. \\

To overcome the above shortcomings of existing applications of deep learning on astronomical vision tasks, especially on galaxy image processing tasks, we have developed a comprehensive framework containing a foundation model, multiple machine learning models for downstream tasks, and a human-in-the-loop (HITL) interface. The foundation model is based on the Swin-Transformer model~\cite{liu2021swin}, and the galaxy images from ssl-legacysurvey project~\cite{Stein2021arXiv} are selected as pre-training data, which contains 76 million galaxy images pulled from the Dark Energy Spectroscopic Instrument (DESI) Legacy Survey~\cite{dey2019overview} Data Release 9. Covering 14,000 square degrees of extragalactic sky in three optical bands (g,r,z), these data are believed to compose a relatively complete description of galaxies in the nearby universe. Different neural networks are then attached to the trained LVM for downstream tasks, including classification, image restoration, outlier detection, etc. The model requires far fewer training samples than the current supervised learning algorithms and is suitable for various purposes. For further performance enhancement, a HITL module based on the FLASK web framework~\cite{grinberg2018flask} is connected to our framework, which takes advantage of human knowledge to further decreases the workload of data labeling and improve the reliability, universality, and interpretability of the framework in different image processing tasks. \\

\section{A foundational Vision Model for Astronomy}\label{subseclargemodel}
Regarding the foundational model, according to deep learning theory~\cite{lecun2015deep}, there has been a proliferation of neural networks featuring progressively deeper architectures, from millions to billions of parameters, that encode prior knowledge about specific domains of problems, such as large language models (LLM)~\cite{Devlin2018bert,dai2019transformer,brown2020language,touvron2023llama} and large vision models (LVM)~\cite{kirillov2023segment,Lanusse2023}. These so-called large models can be used as the backbone for various tasks, offering themselves proficient few-shot learners capable of handling various data-processing challenges. In this context, the LVM is developed on the basis of the Swin-Transformer architecture. The LVM is trained in an unsupervised manner using 76 million stamp images with ${g, r, z}-$bands~\cite{Stein2021arXiv,Stein2022} from DESI Legacy Imaging Surveys. More details on the LVM are presented in the remaining parts of this section.\\ 

\subsection{Design of the Large Vision Model}
\begin{figure}[htb]
  \centering
     \includegraphics[width=1\textwidth]{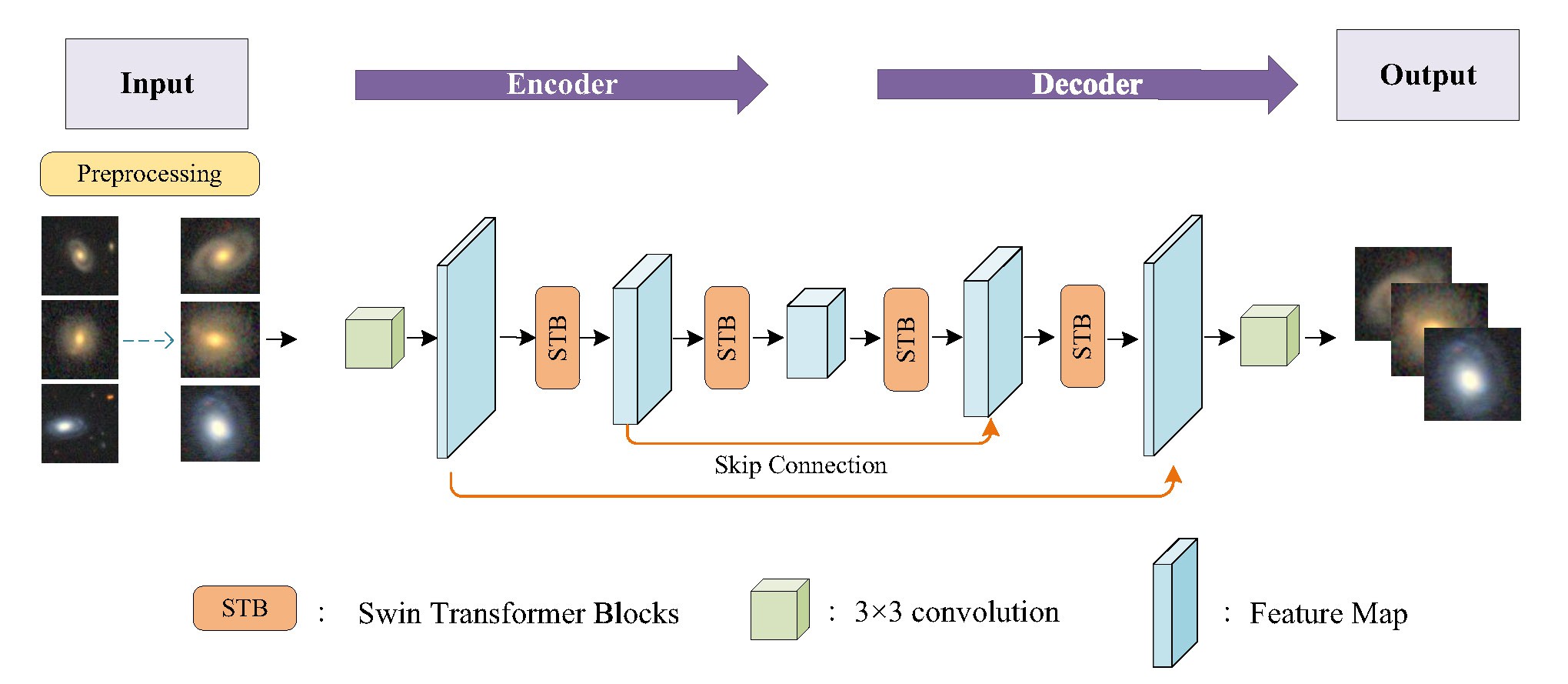}
   \caption{\textbf{The Structure of the Large Vision Model (LVM).}}
   \label{model_structure}
\end{figure}

Figure~\ref{model_structure} illustrates the architecture of our LVM, which is based on the SUNET framework~\cite{fan2022sunet} and boasts a parameter count of approximately 100 million. For demonstration, Figure~\ref{model_structure} only displays 4 layers of the Swin-Transformer Block (STB). The core structure of the LVM follows an encoder-decoder paradigm, with Swin-Transformers serving as the fundamental building blocks. The use of Swin-Transformers is pivotal in amplifying the interpretability of abstract image features, a crucial factor for grasping the fundamental elements and inherent characteristics residing in the data. Note that Swin-Transformer effectively processes local information through its window attention and gradually expands the receptive field and integrates global information through shifted window attention. Besides, compared to traditional Transformers or ViTs, Swin-Transformer significantly reduces computational complexity and memory requirements while maintaining model performance. In essence, LVM attempts to reconstruct the original images utilizing the sparse features extracted by the encoder and decoder~\cite{chan2022redunet}. This process involves learning a mapping function that translates 3 channels of 2-dimensional image data into semantic features in the latent space. \\

\begin{figure}
  \centering
    \includegraphics[width=0.9\textwidth]{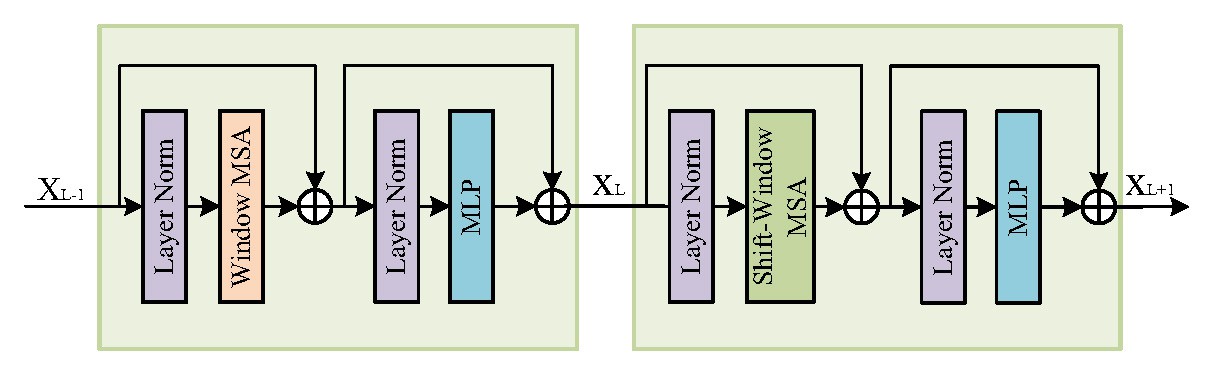}
  \caption{ \textbf{The Structure of the Swin-Transformer Layer (STL) in the LVM.}}
  \label{model_STL}
\end{figure}

The LVM encoder comprises four layers, each containing eight consecutive Swin-Transformer layers (STL), and the decoder has an identical structure. $3 \times 3$ convolution kernels are adopted along with the Swin-Transformer Block.  Moreover, by engaging in feature processing across multiple tiers and assimilating global information, the model attains a deeper understanding of the interconnectedness and dependencies inherent within these images, which, in turn, facilitates more effective inference processes. Notably, this architectural design, featuring a deep feature pyramid, significantly fortifies the performance of the model across tasks encompassing various scales. Figure~\ref{model_STL} shows two STLs. These blocks encompass normalization layers, a window-based self-attention layer (Window MSA), a shift window-based self-attention layer (Shift-Window MSA), and a multilayer perceptron (MLP). These layers enhance the perceptive capabilities of the Swin-Transformer compared to traditional convolutional neural networks. These components are integrated into the U-net structure, effectively increasing the receptive field of the neural network, which substantially amplifies the capacity of the model for representing data samples within the feature space.\\

\subsection{Pre-training of the Large Vision Model}
\label{pretrainsec}
The LVM undergoes pre-training through a self-supervised method~\cite{he2021masked,xie2022simmim} using celestial object images from the DESI Legacy Imaging Surveys DR9. Each instance presented to the model is a galaxy image, generating an identical galaxy image as its output. These images comprise three channels denoting the $[g, r, z]$-bands and are resized to $152 \times 152 \times 3$ pixels. The LVM initially compresses galaxy images into feature vectors via the encoder and subsequently reconstructs galaxy images based on these vectors. By utilizing the Mean Squared Error (MSE) loss, the difference between the reconstructed galaxy images and the originals can be measured, which fosters effective learning of galaxy image representations.\\

When dealing with galaxy images, there is a challenge from the varying effective sizes of different galaxy images. Leaving this problem unsolved could lead to some galaxies appearing relatively small in the images, making effective analysis and recognition problematic. To overcome this problem, an OpenCV-based algorithm is devised to adaptively cropping images~\cite{opencvlibrary}. The algorithm calculates the effective area that galaxies occupy at each stamp according to the grayscale level. Then, it cuts and resizes the original images to create new stamp images with a fixed size of $128 \times 128 \times 3$. This step ensures that each galaxy occupies an appropriate area within the image without losing much information, thus easing subsequent processing and analysis. In addition, data augmentation are performed by applying flips, rotations, and croppings to generate a more diverse set of training samples from the images to enhance the coverage of the training sets in latent space, which enhances the generality of the model and its overall robustness. \\

The batch size in the pre-training stage is set to 512 by balancing the efficiency and hardware limits. In each iteration, the MSE is computed for all images in a batch and subsequently updated the model parameters using the Adam optimizer. It takes approximately 196 hours using 8 NVIDIA A100s with 80GB of graphic memory to train the LVM. After training, the encoder within the LVM acquires the ability to learn the features inherent to celestial objects. It is feasible to cut this encoder from the LVM and connect it to the following neural networks for further training. This extended training could involve various downstream tasks and the HITL strategy, which will be elaborated on in the following section. \\

\section{Multiple Downstream Tasks Training for the Large Vision Model}\label{section3}

\subsection{Training of Multiple Downstream Tasks}

\begin{figure}
\centering
\includegraphics[width=0.95\textwidth]{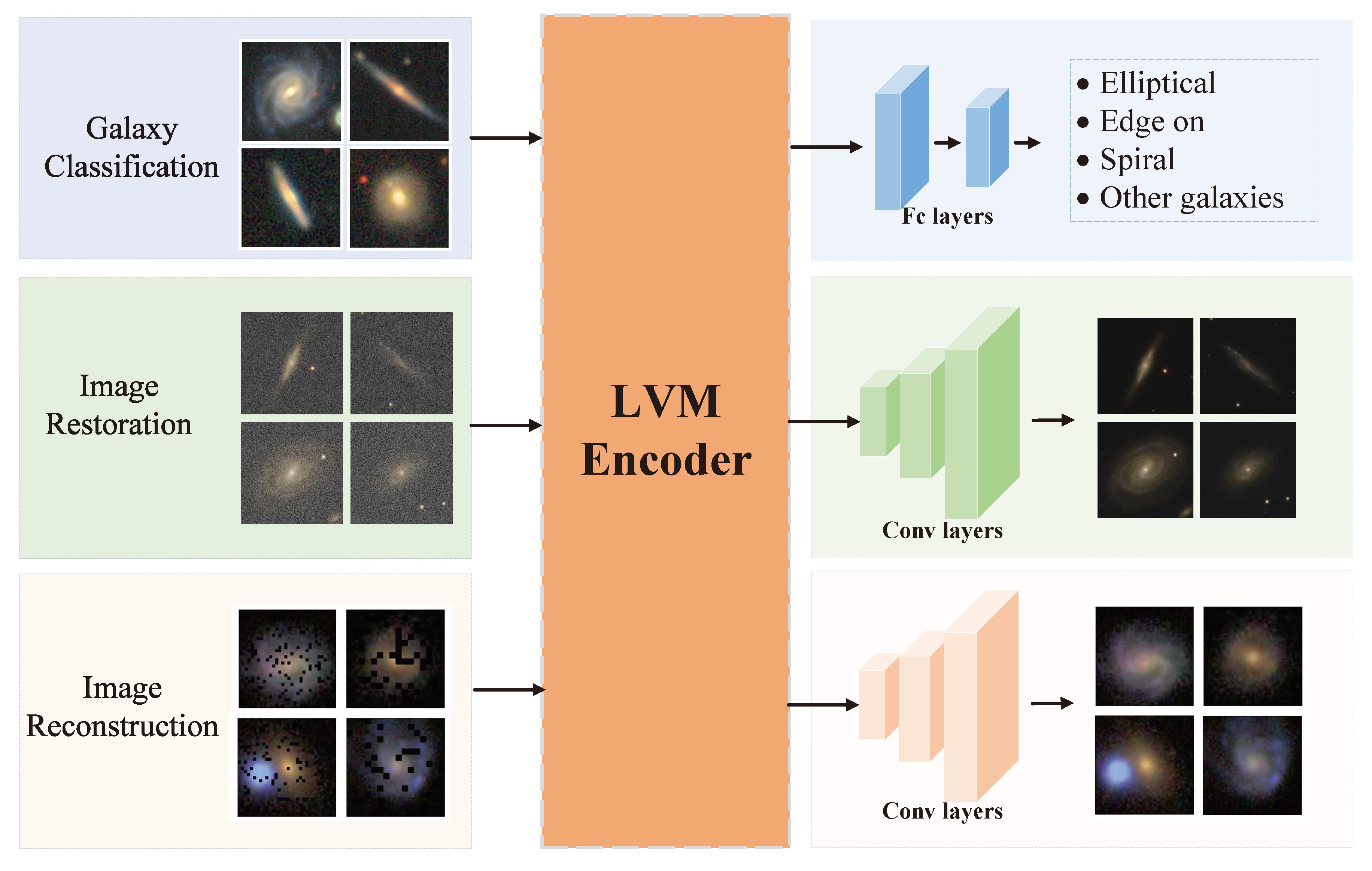}
\caption{ \textbf{The schematic draw of multiple task training process.}}
\label{mask_task}
\end{figure}

Given that common foundational knowledge is suitable for various downstream tasks, there is a possibility of enhancing the encoder's proficiency within the LVM by concurrently engaging in multiple downstream tasks. This approach aims not only to enhance the versatility of the LVM, but also to optimize task-specific performance. In line with this philosophy, three downstream tasks have been identified: galaxy classification, image restoration, and image reconstruction. For each task, a task-specific neural network is incorporated alongside the LVM encoder, as illustrated in Figure~\ref{mask_task}. During the multitask training stage, updating the model parameters for the entire multitask framework with these Tasks. An active learning strategy is used to dynamically adjust the proportion of training dedicated to different tasks.\\

The image classification task aims to classify galaxies according to their morphologies. This is achieved by adding two fully connected layers following the LVM encoder. Additionally, a dataset containing images of galaxies has been constructed, with four distinct classes: Elliptical, Edge-on, Spiral, and Other galaxies (including irregular and merging galaxies). These Galaxy images are obtained from the DESI Legacy Imaging Surveys, while the labels indicated their morphologies are obtained from the Galaxy Zoo 2 project \footnote{\url{https://data.galaxyzoo.org/}}. The data processing method discussed in~\cite{zhu2019galaxy} is used to obtain high-quality labels. For the multitask training, a training dataset containing 500 galaxy images per category, totaling 2000 images, is used for model training. Furthermore, a test dataset, consisting of 250 images in each category, totaling 1000 images, is utilized to assess the model's performance.\\

The image restoration task strives to generate high-quality original images from blurred ones. This is achieved by incorporating a decoder module with convolutional layers following the LVM. The dataset comprises two components: 1) reference images are high-quality raw galaxy images obtained from the DESI Legacy Imaging Surveys ; 2) blurred images are generated by introducing noise and blurred PSFs using the method outlined in~\cite{schawinski2017generative}. In the experiment, the Moffat model is employed, assuming that the PSFs with a full width at half-maximum (FWHM) are distributed in the 2.0 to 8.0 pixels range. Additionally, a Gaussian function is assumed as the noise source, with a standard deviation uniformly distributed between 1.0 and 15.0, to simulate blurred data. These blurred images simulate the degradation and noise processes in real observations. For the purpose of multitask training, a training dataset containing 1000 blurred images and a test dataset consisting of 100 images are utilized to assess the model's performance. Notably, both the training and test datasets are derived from simulated data, ensuring a controlled environment for the training and evaluation of the model.\\

The image reconstruction task aspires to mend the obstructed sections of images, facilitating the segmentation of individual galaxies from several adjacent galaxies. This is achieved by integrating a decoder module that comprises convolutional layers following the LVM. The dataset for this task comprises two components: 1) reference images, original images without bad pixels or other defects obtained from the DESI Legacy Imaging Surveys; 2) masked images, original images masked by varying patch sizes, from $0\%$ to $70\%$ with a random scale, which emulates image degradation process that may occur during observation and acquisition. For multitask training, a training dataset consisting of 1000 data pairs and a test dataset containing 100 pairs are employed to assess the model's performance in image reconstruction.\\

Various loss functions are applied during the training process for different tasks. Cross-entropy is the loss function for galaxy classification tasks, while the Mean Squared Error (MSE) is employed for image restoration and reconstruction tasks. We adopt two training strategies for training downstream tasks models for comparative studies. The first strategy ($Multi\_uniform$) maintains equal weights for each task during training, meaning that the training proportion for each task is consistent. The second strategy ($Multi\_active$) actively updates the training proportion of each task according to the characteristics and performance during the training process. Expressly, this strategy quantifies the allocation proportion of each task during training by evaluating its performance (such as MSE or F1 score) on the test set. Tasks that demonstrate better performance metrics are allocated a smaller proportion of training data, while tasks with lower performance metrics receive a larger proportion.


\subsection{Performance Evaluation}
A series of comparative experiments are conducted to evaluate the performance of an untrained multitasking model ($Pre\_train$ model) against a multitasking-trained model ($Multi\_uniform$ model and $Multi\_activate$ model) using the training strategies mentioned earlier across image classification, image reconstruction, and image restoration tasks. Note that, the $Pre\_train$ model is given by using the frozen training approach, i.e., the weights of the LVM are kept constant and only the weights in the task head is updated.\\

In the image classification task, our model was trained using a dataset of 1000 images and its performance was evaluated on a separate set of 500 images. The results are presented in Table~\ref{classresultqq}, showcasing the classification accuracy (Acc), precision (Pre), recall (Recall), and F1 scores in three distinct training strategies. Our analysis indicates that models trained on multiple tasks outperform those exclusively trained on the restoration task ($Pre\_train$ model) with respect to classification accuracy and various metrics. In particular, the actively selected task strategy ($Multi\_activate$ model) demonstrates a substantial improvement in accuracy compared to the other two. These results suggest that the $Multi\_activate$ training strategy can augment the model's classification performance, and actively selecting the task further enhances accuracy by sacrificing precision, Recall, and F1 slightly (see Table~\ref{classresultqq}).\\

\begin{table*}[htb]
\centering
\begin{tabular}{ccccc}
\hline
\hline
{Classification Task} & Acc & Pre & Recall & F1 \\
\hline
Pre\_train model& 0.784 & 0.767 & 0.794 & 0.771 \\
\hline
Multi\_uniform model& 0.842 & 0.844 & 0.850 & 0.846 \\
\hline
Multi\_activate model& 0.854 & 0.823 & 0.834 & 0.844 \\
\hline
\hline
\end{tabular}
\caption{The classification results of model trained with various training strategies.}
\label{classresultqq}

\end{table*}

Regarding the task of image restoration, to gauge the effectiveness of our model, we employ a variety of metrics, including PSNR (Peak Signal-to-Noise Ratio), MSE (Mean Squared Error), and SSIM (Structural Similarity Index). These metrics are utilized to evaluate the agreements between the processed images and the original unprocessed images. Higher PSNR, higher SSIM, and lower MSE indicate better agreements. Our findings, as presented in Table~\ref{restore_result}, indicate that the models trained with the $Multi\_uniform$ strategy outperform the $Pre\_train$ ones, and the multitasking plus active learning strategy ($Multi\_activate$) brings the optimal model.\\

\begin{table*}[htb]
\centering
\begin{tabular}{cccc}
\hline
\hline
{Restoration Task} & MSE & PSNR & SSIM \\
\hline
Blurred images & 0.00094 & 31.11 & 0.48 \\
\hline
Pre\_train model& 0.00084 & 31.31 & 0.51 \\
\hline
Multi\_uniform model& 0.00083 & 31.35 & 0.54 \\
\hline
Multi\_activate model& 0.00049 & 33.34 & 0.56 \\
\hline
\hline
\end{tabular}
\caption{Image restoration results under different training strategies.}
\label{restore_result}
\end{table*}

\begin{figure}
  \centering
    \includegraphics[width=0.9\textwidth]{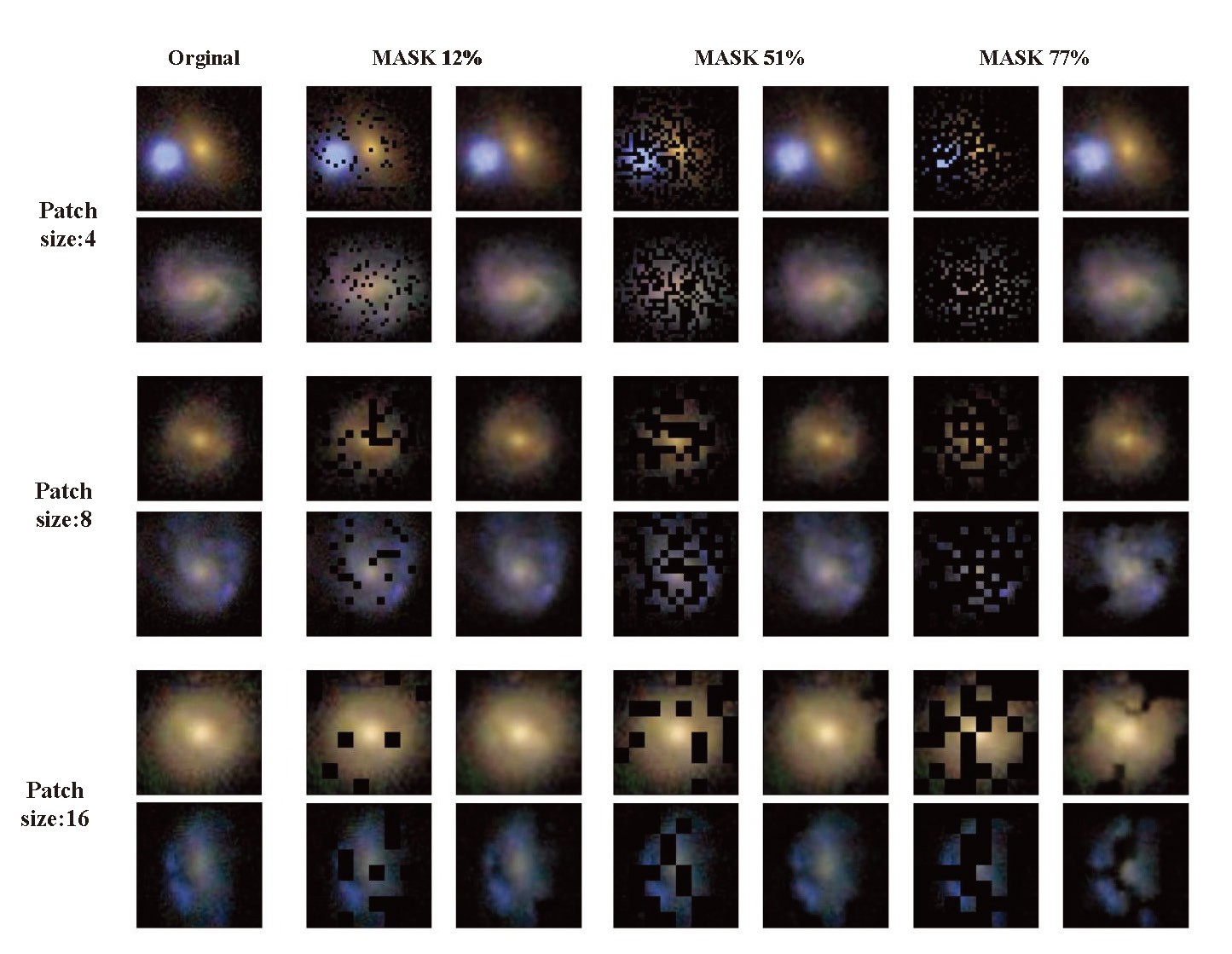}
  \caption{\textbf{Images Reconstructed by the Model with Varying Patch Block Sizes and Masking Proportions}}
  \label{mask_result}
\end{figure}

For the image reconstruction task, a dataset consisting of 1000 samples was used for training and 100 for testing. The performance was also evaluated using metrics of PSNR, MSE, and SSIM. Multitasking-trained models ($Multi\_uniform$ and $Multi\_activate$ models) demonstrate superior performance in the image reconstruction task, surpassing the non-multitask-trained model ($Pre\_train$ model) in the reconstruction of masked regions, as shown in Table~\ref{maskresult}. Furthermore, a comprehensive evaluation of the outcomes was conducted by analyzing the image reconstruction performance under varying levels of missing data and patch sizes. The results presented in Table~\ref{maskresult} and Figure~\ref{mask_result} demonstrate the remarkable performance of the model, even when dealing with highly degraded images with a content missing rate of up to 70$\%$ and a patch size of 8×8 (for input data of size 128×128).

\begin{table*}[htb]
\centering
\begin{tabular}{cccc}
\hline
\hline
{Reconstruction Task}  & MSE & PSNR & SSIM \\
\hline
Masked images & 0.0248 & 15.64 & 0.36 \\
\hline
Pre\_train model& 0.0089 & 22.36 & 0.49 \\
\hline
Multi\_uniform model& 0.0040 & 26.07 & 0.61 \\
\hline
Multi\_activate model& 0.0038 & 26.84 & 0.64 \\
\hline
\hline
\end{tabular}
\caption{Image reconstruction results with models trained with different strategies.}
\label{maskresult}
\end{table*}

\begin{table*}
  \centering
\begin{tabular}{cccc}
\hline
\hline
Patch size & MSE & PSNR & SSIM  \\
\hline
$4\times4$ (Masked images) & 0.03187 & 15.92 & 0.27  \\
\hline
Multi\_activate model & 0.0067 & 26.54 & 0.58 \\
\hline
$8\times8$ (Masked images) & 0.023 & 17.55 & 0.36  \\
\hline
Multi\_activate model & 0.0049 & 24.81 & 0.55  \\
\hline
$16\times16$ (Masked images) & 0.03149 & 17.47 & 0.42  \\
\hline
Multi\_activate model & 0.0112 & 23.61 & 0.48  \\
\hline
\hline
\end{tabular}
\caption{Statistical Analysis of Image Reconstruction Performance. Comparison of PSNR, SSIM, and MSE under varying patch sizes and masking proportions (Using Frozen and Fine-tuned LVM Model Parameters).}
\label{difflevelblock}
\end{table*}

Above all, multitasking training in conjunction with active learning significantly enhances the performance of the model across different tasks. In comparison to the model that has not undergone multitask training, the utilization of multitask training facilitates a more effective acquisition of feature representation and enhances the generalization ability of the neural network, rendering it suitable for a variety of astronomical image processing tasks. \\

\section{Deployments of two sample Applications}\label{section4}

Two astronomical vision tasks are chosen to showcase the capabilities of our LVM model: galaxy morphology classification and strong lens finding from a large field of view. Here, the LVM is utilized as the backbone and two separate downstream models for the two tasks. Detailed information on each of these applications is presented below.\\

\subsection{Classifying Galaxy Morphology with Few-Shot Learning based on LVM} \label{AP:fewshot}
\begin{figure}
  \centering
    \includegraphics[width=\textwidth]{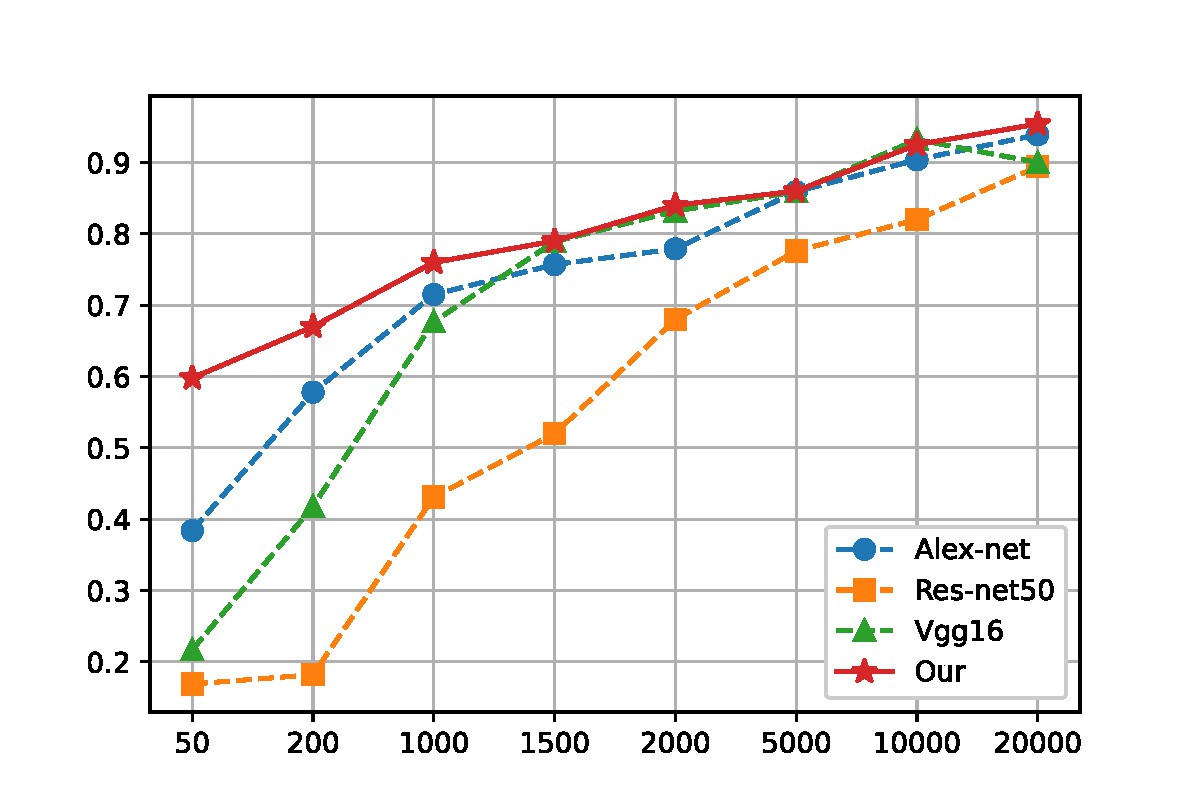}
  \caption{\textbf{Classification Accuracy of Different Models on Dataset with Different Sizes.} This figure shows the classification accuracy of four models (Our Model, AlexNet, VGG16, and ResNet50) on the galaxy classification task under data with different sizes. }
  \label{class_result}
\end{figure}

To further evaluate the efficacy of the proposed algorithm, its performance in the morphological classification of galaxies within the DESI Legacy Imaging Surveys using the few-shot learning approach has been assessed. The training and testing sets include image data from the DESI Legacy Imaging Survey and labeled from the Galaxy Zoo project. The galaxies are categorized into five types~\cite{walmsley2022galaxy}. After the LVM encoder, a fully connected neural network is involved for the task of galaxy morphological classification and then trained with the above training sets. A comparative analysis is performed by evaluating the outcomes of our model and those of AlexNet~\cite{krizhevsky2017imagenet}, VGG16~\cite{simonyan2014very}, and ResNet50~\cite{he2016deep}, which are deep learning architectures those have been proven effective in various image recognition tasks.  As Figure~\ref{class_result} illustrates, the LVM + Downstream Tasks model maintains higher accuracy, especially in scenarios with minimal data (only ten images per class). Moreover, as the amount of data increases, the model performance gradually improves, further confirming its scalability on large datasets. These experimental results not only demonstrate the effectiveness of the LVM + Downstream Tasks model in galaxy morphology classification tasks, but also reveal its stability and generalization ability when dealing with datasets of different sizes. 


\subsection{Identifying Strong Lensing systems with the LVM + Mask R-CNN}\label{AP:Detection}
\begin{figure*}
  \centering
    \includegraphics[width=0.98\textwidth]{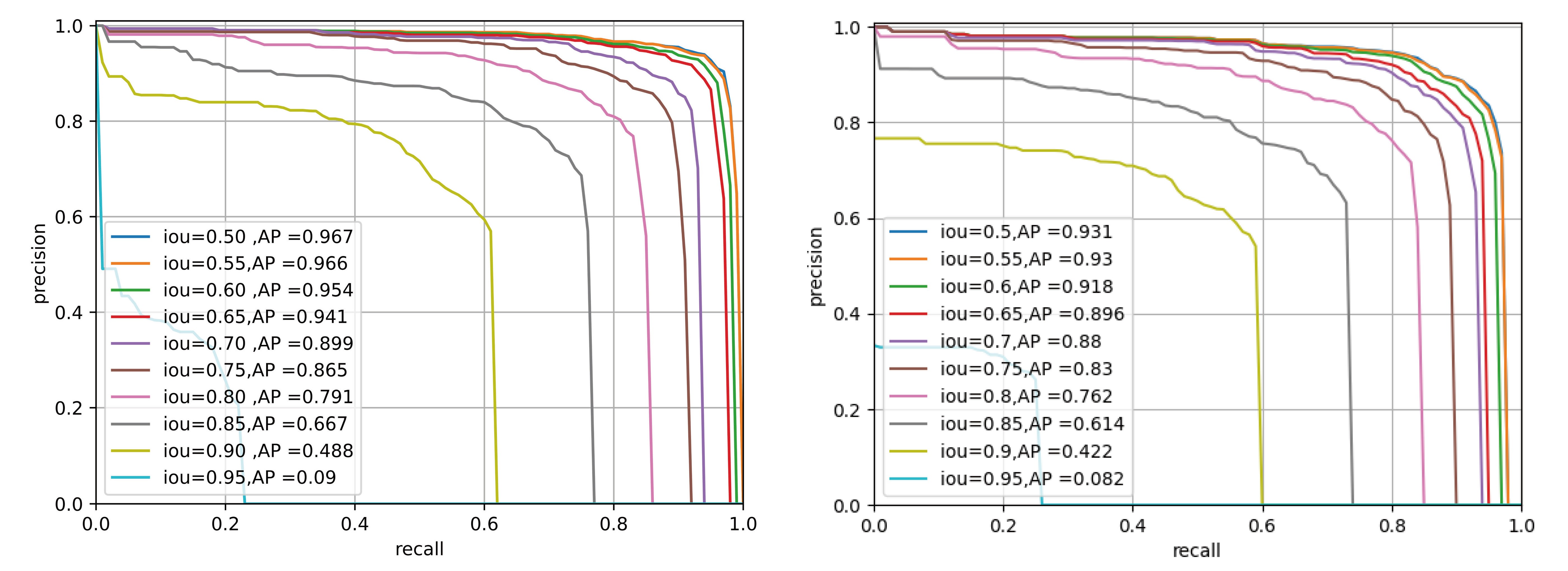}
  \caption{\textbf{Target detection results with different backbone.} In the left panel, it is the target detection outcomes achieved using Mask R-CNN with the LVM as the backbone. Meanwhile, the right panel displays the target detection results obtained with Mask R-CNN using ResNet50 as the backbone. As evident in these figures, the LVM significantly enhances the detection capabilities of the neural network.}
  \label{detection_result0}
\end{figure*}

To replicate this trend in source detection, a strong lens dataset is constructed firstly, which comprises 1000 training images and 1000 testing images. These images were extracted from the DESI website \footnote{\url{https://www.legacysurvey.org/}} using the catalog of strong lensing system candidates available in the \textit{NeuraLens Database}\footnote{\url{https://sites.google.com/usfca.edu/neuralens/}}. For the downstream task of finding strong lensing systems within a large field of view, Mask R-CNN ~\cite{he2017mask} was chosen as our model. Additionally, ResNet50 was employed as the backbone of Mask R-CNN for comparison. The results, presented in Figure~\ref{detection_result0}, demonstrate that our LVM + Mask R-CNN model achieved an impressive average precision (AP) of 96.7$\%$ with 1000 training images. In contrast, the ResNet + Mask R-CNN model achieved a slightly lower AP value of 93.1$\%$. This comparison underscores the effectiveness of our LVM approach in enhancing the performance of Mask R-CNN for strong lens detection.

\section{Large Vision Model with the Human-in-the-loop}\label{section5}
\begin{figure}
  \centering
    \includegraphics[width=1\textwidth]{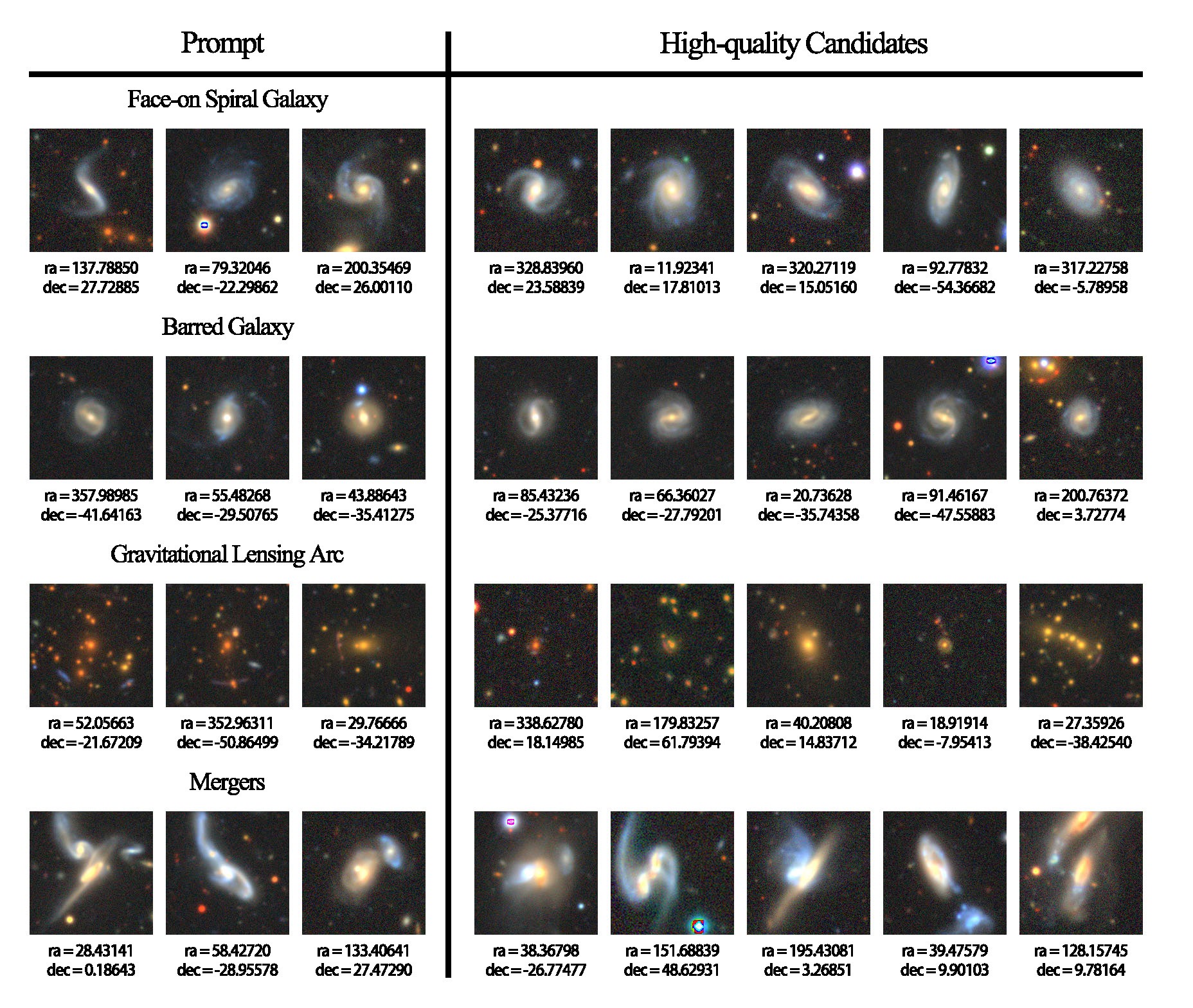}
  \caption{ \textbf{The input prompts and recommended objects after several rounds of interaction.} For simple tasks like face-on spirals, one round is enough, and for more complex tasks like mergers, no more than 10 rounds are used.}
  \label{HITL_InterestedFound}
\end{figure}

To interactively integrate human knowledge, we develop a HITL module~\cite{Wu_2022} based on the Flask Web Framework \footnote{\url{https://flask.palletsprojects.com/}} and integrate it into the LVM. Taking the binary classification task as an example, a Multilayer Perceptron (MLP)~\cite{6302929} model with a hidden layer size of 2048 is used to predict the types of galaxies and introduce the HITL module with an adaptive algorithm to find potential objects to boost its purity, completeness, or some other measurements. These objects will be labeled and included in the training sets in the MLP training procedure. With this module, astronomers can create training sets iteratively from scratch for their specific purposes and direct the AI model's optimization path as necessary.\\

to evaluate its feasibility, the HITL is used to identify spiral galaxies from ellipticals. It achieves an Area Under the Precision-Recall Curve (AUPR) of 0.8895 by starting with 10 initial prompts (5 positive and 5 negative) and following one interaction step with 10 recommended examples. The performance surpasses that of training the LVM with 30 examples (15 positive and 15 negative, AUPR = 0.8683) and training a ResNet18 with 100 examples (50 positive and 50 negative, AUPR = 0.8561). It is comparable to training the LVM with 1000 random examples (AUPR = 0.8921). Figure~\ref{HITL_InterestedFound} presents the tests of the HITL on specific target identification tasks with few-shot learning, such as finding galaxies with bars, strong lensing systems, and galaxy mergers, which further proves the capacity of this module and shows the broad application potential for different tasks. More details are discussed below.\\

\subsection{Design of the Human-In-The-Loop Module}
\begin{figure}
  \centering
    \includegraphics[width=0.9\textwidth]{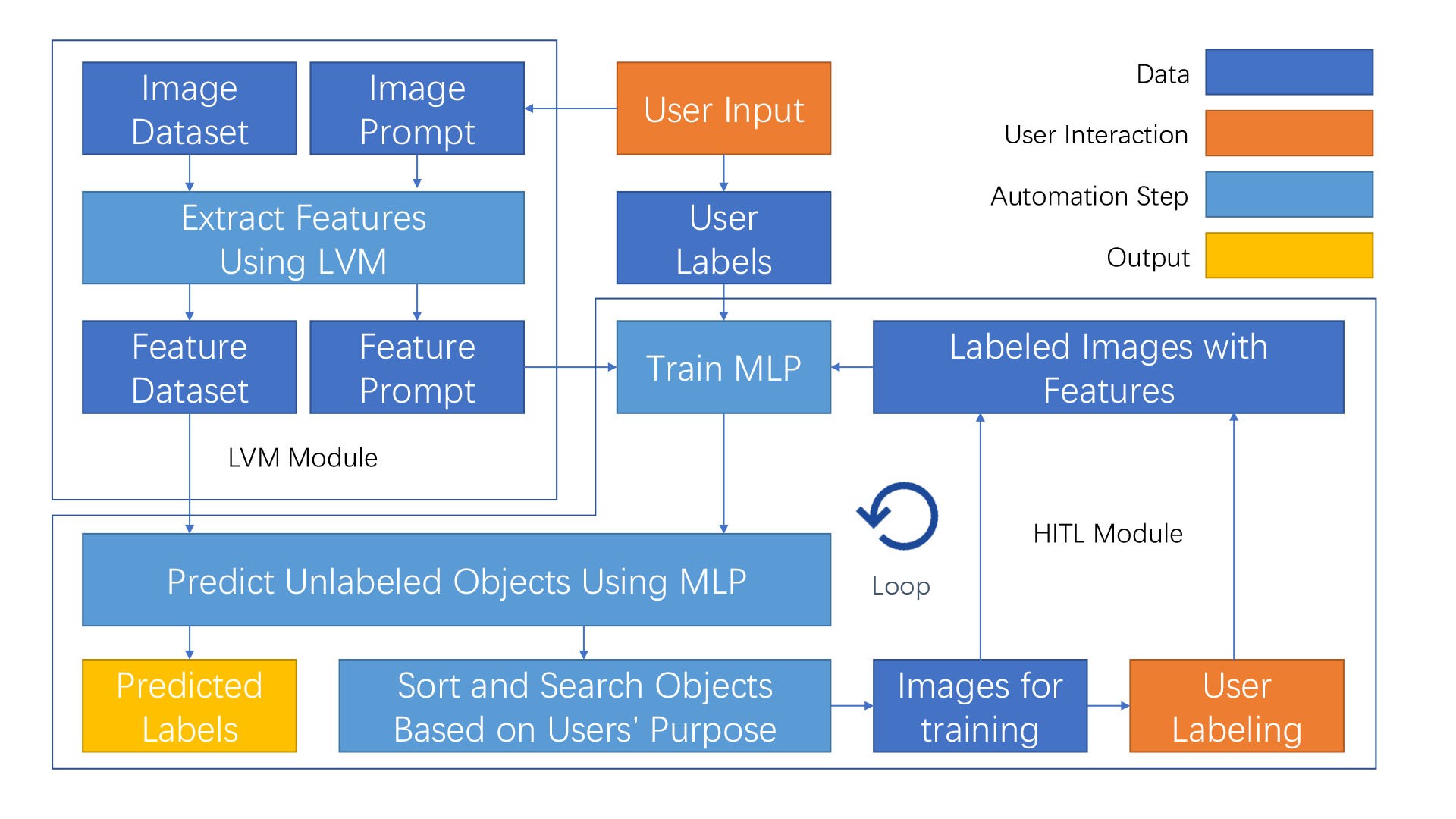}
  \caption{\textbf{The overall design of the human-in-the-loop (HITL) module}}
  \label{HITL_Module}
\end{figure}

\begin{figure*}
  \centering
    \includegraphics[width=0.9\textwidth]{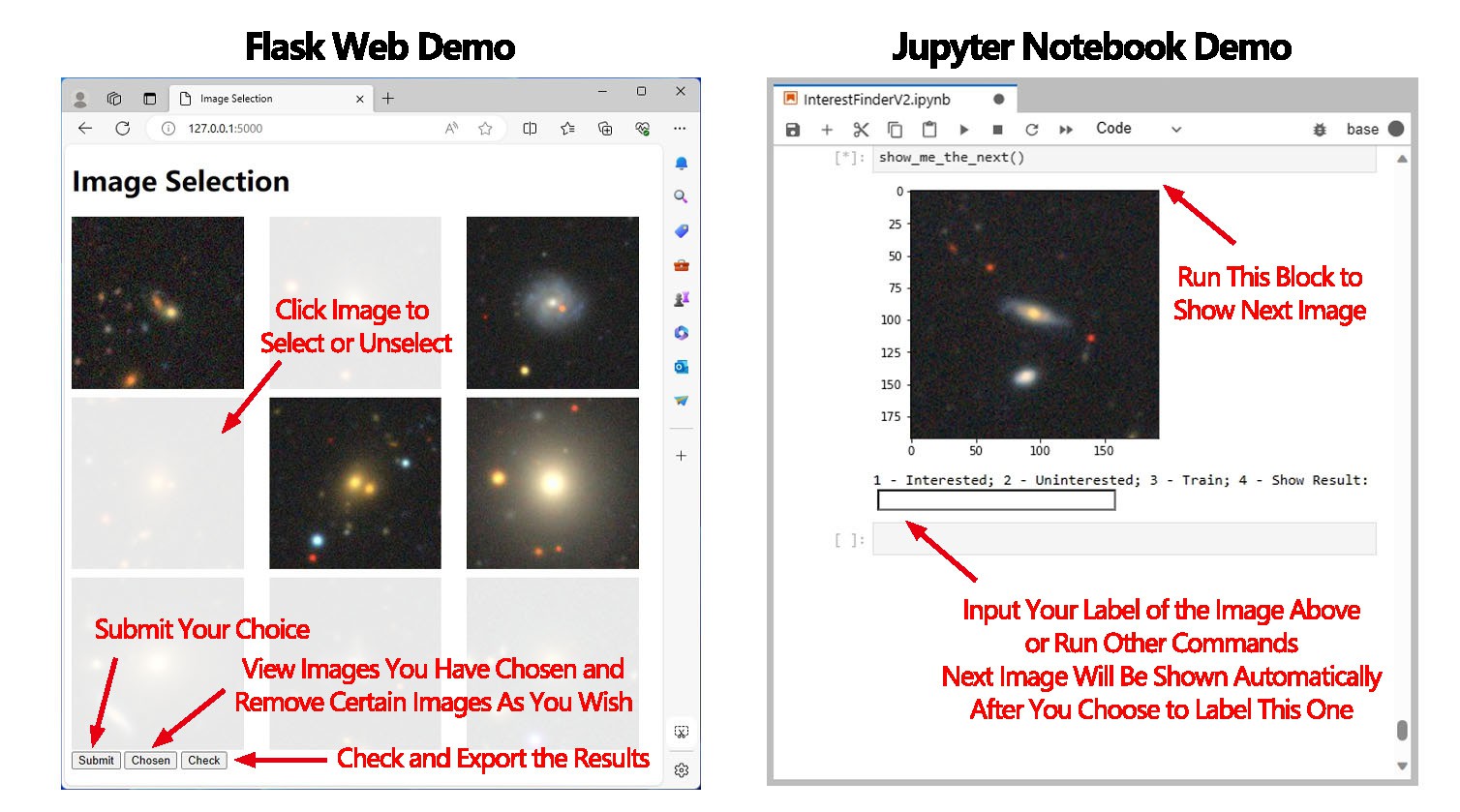}
  \caption{ \textbf{ The interface of our HITL classification web demo and Jupyter notebook demo.} In the web demo, Users can classify images by clicks and view classification results as well as export them as reference for further applications. In the Jupyter notebook demo, users can do the same work by view images and input their commands.}
  \label{HITL_Module_webdemo}
\end{figure*}

The overall design of the human-in-the-loop (HITL) module is shown in Figure~\ref{HITL_Module}, which contains a foundational LVM model, a downstream network, and a HITL module. The HITL module contains a frontend and a backend. With the frontend constructed with HTML and CSS, the image can be labeled to train the AI model by clicking on images (Left Panel in Figure~\ref{HITL_Module_webdemo}). Then, these actions will be passed to the backend constructed with the Flask framework, which communicates between the HITL module and LVM. In addition, a user interface based on the Jupyter Notebook is also available for this purpose for those who do not run web applications (Right Panel in Figure~\ref{HITL_Module_webdemo}).

The downstream network is an MLP with a hidden layer of size 2048 running on fixed features extracted by our LVM, which costs significantly less computing resources than training a model for a specific task from scratch but proved adequate to bring the ability of our LVM into full play. To make the interaction beneficial to the model for various purposes, we set a parameter $0 \leq \alpha \leq 1$, a threshold of the ratio between positives and negatives labeled during the latest interaction loop ($P/N$). When $P/N \leq \alpha$, the HITL module will select objects with higher scores for user labeling; when $P/N > \alpha$, the HITL module will select objects with lower scores for user labeling. Changing $\alpha$ can guide the downstream model to converge in the required directions. For example, $\alpha=0.9$ and $\alpha=0.1$ give models for high precision and recall rates, respectively, while $\alpha=0.5$ leads to a model with high AUC and AUPR.\\

\subsection{Comparisons to the models of conventionally supervised learning}

To evaluate the effectiveness of our HITL module, the classification results of 8,522 galaxy images from Galaxy Zoo DECaLS~\cite{2022MNRAS.509.3966W} by using different approaches are compared, which are not included in the training sets used to train our LVM, with LVM + HITL and traditional supervised learning approaches. These galaxies are classified into Face-on Spirals (4546 images) or others (3976 images, including ellipticals and edge-on galaxies) ~\cite{2022RAA....22e5002Z}. We have tested the performance of supervised ResNet18\footnote{\url{https://pytorch.org/vision/stable/index.html}}, our LVM, and our LVM + HITL with different sizes of training datasets. \\

\begin{figure*}
  \centering
  \includegraphics[width=0.9\textwidth]{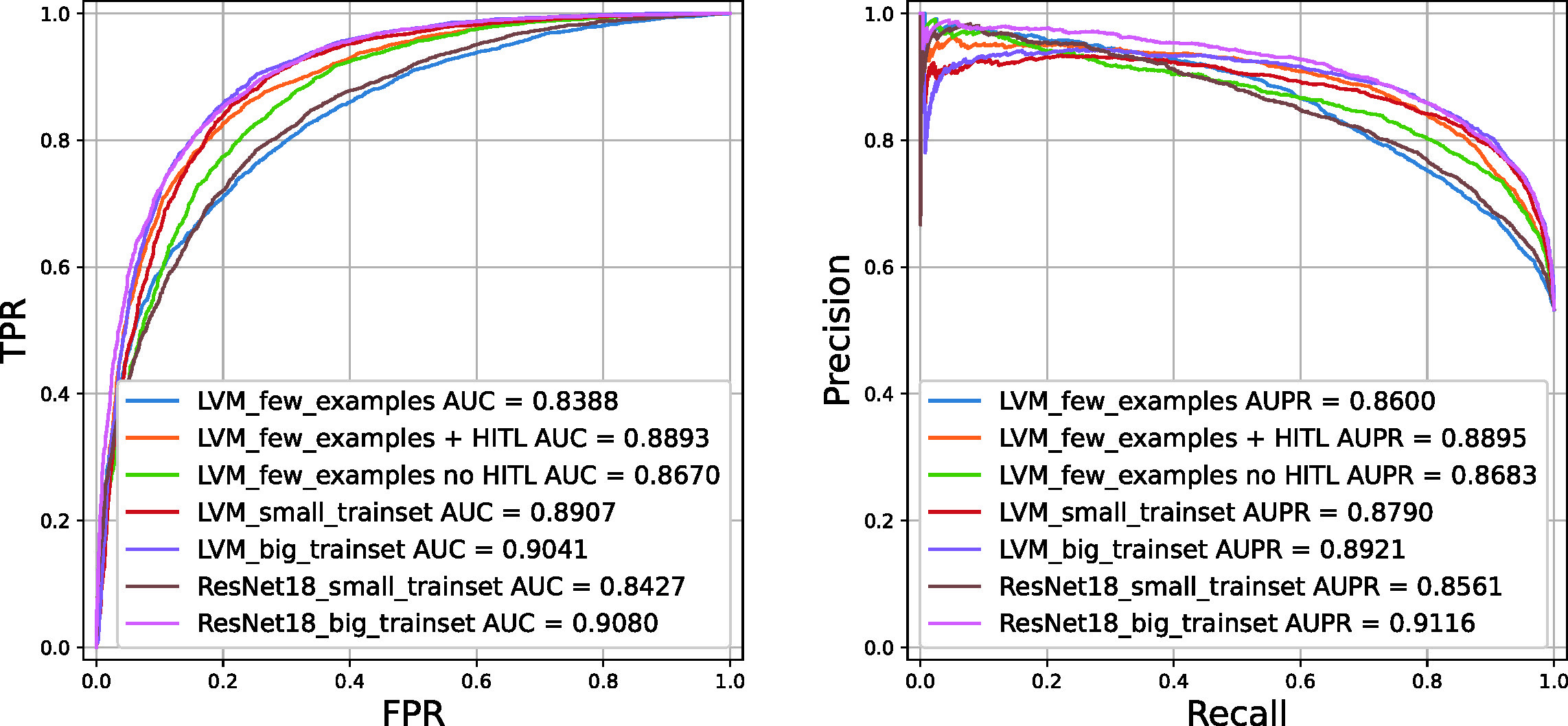}
  \caption{ \textbf{ Comparison of the performance of ResNet18, our LVM and our LVM + HITL under training set with the same number of images.}}
  \label{HITL_vs_Supervised}
\end{figure*}

Specifically, we first give five positives and five negatives to the downstream classification network upon the LVM, and the outcomes are named $LVM\_few\_examples$ in Figure~\ref{HITL_vs_Supervised}. Then, we label an extra ten examples recommended by the HITL module and see a boost in performance ($LVM\_few\_examples + HITL$ in Figure~\ref{HITL_vs_Supervised}). For comparisons, 15 positive and 15 negative examples are given to our LVM ($LVM\_few\_examples$ without HITL" in Figure~\ref{HITL_vs_Supervised}), 50 positive and 50 negative examples to our LVM and ResNet18 as $small\_training\_set$ in Figure~\ref{HITL_vs_Supervised}, and 1000 randomly chosen examples as $big\_training\_set$ in Figure~\ref{HITL_vs_Supervised}. As shown, the LVM + HITL approach performs better than LVM alone, equivalent to the methods with $big\_training\_set$. \\

\subsection{Desired Targets finding from BGS with LVM + HITL}

To test the feasibility of the LVM + HITL approach on real observation, we constructed serial tasks for desired object detection from 201,319 galaxy images selected from the DESI Bright Galaxy Survey (BGS)~\cite{2023AJ....165..253H}, which is a selection of bright galaxy objects in the DESI Legacy Imaging Surveys. These galaxies are excluded from the training sets for LVM. These galaxies have half-light radii between 6.4 and 9.6 arcsec. These objects are cropped from fits files of g, r, and z channels into images of size $(H, W, C) = (192, 192, 3)$ with a pixel scale of 0.262 arcsec/pix to maintain the surroundings, which could be beneficial for identifying objects such as gravitational lensing arcs and mergers.\\

Firstly, two characteristic types of galaxies are selected for target-finding experiments: face-on spiral galaxies and barred galaxies, representing relatively common objects. Starting with only five positives, our LVM + HITL acquires precision rates of 0.91 and 0.75 for identifying face-on spiral galaxies and barred galaxies, respectively, within ten rounds of interactions. The findings demonstrate that our LVM + HITL method can help astronomers identify their desired targets for specific scientific goals with a reference sample containing a few objects.

Moreover, the LVM plus HITL model are utilized to identify strong gravitational lensing systems and galaxy mergers to study its feasibility in searching for rare and complex astronomical objects. Figure~\ref{HITL_InterestedFound} shows that our approach can discover these targets successfully. However, the outcomes include many more false positives than the tasks for identifying common objects, e.g.,  the precision rate of finding galaxy mergers is only 0.15. In principle, this issue can be improved by adopting an appropriate way for LVM to deal with the feedback from the HITL module beyond depending on $\alpha$ alone, which is a primary part of our future explorations.

\section{Summary}

Above all, we have created a framework that utilizes a HITL module on top of an LVM for various astronomical vision tasks. Expressly, the downstream neural networks, combined with the LVM, allow for versatility without the need for expensive re-training, and the HITL module incorporates human knowledge to guide the AI model to regress towards specific objectives, which reduces the workload for composing training sets and enhances the framework's universality and interpretability. The experiments have shown that our framework outperforms traditional supervised machine learning models in classical vision tasks in Astronomy, such as astronomical object detection, galaxy morphology classification, and observational image reconstruction, etc.  Considering that the reliability of AI models for handling scientific data is crucial for valid discoveries ~\cite{HCM2023,Phys2024533,PhysRevLett.131.160202}, we have evaluated our framework's reliability in different experiments with the labels in the datasets of Galaxy Zoo 2. However, for the data in the bands beyond g,r,z and those given by space-borne telescopes, Galaxy Zoo 2 is insufficient. Therefore, to assess the framework's reliability in a broader context in the future, we are constructing a standard dataset for galaxy images covering larger feature space with various observations. Moving forward, with the transfer learning strategy, we plan to extend the framework to encompass various data modalities, including photometry, spectrum, and lightcurves. It will lead to a continually evolving AI model that can proficiently handle intricate datasets from various projects, such as DESI, LSST, Euclid, and CSST, which is crucial in the age of multimessage astronomy.\\


\section*{Acknowledgments}
The authors acknowledge the support from the National Natural Science Foundation of China with funding numbers 12173027, 12303105 and 12173062, National Key R \& D Program of China (No. 2023YFF0725300, 2022YFF0503402), science research grants from the Square Kilometre Array (SKA) Project with NO.2020SKA0110100, the Science Research Grants from the China Manned Space Project (No. CMS-CSST-2021-A01, CMS-CSST-2021-A07, CMS-CSST-2021-B05) and the CAS Project for Young Scientists in Basic Research (No. YSBR-062). This work was supported by the Young Data Scientist Project of the National Astronomical Data Center and the Program of Science and Education Integration at the School of Astronomy and Space Science, University of Chinese Academy of Sciences. The authors also thank George Stein for publicly sharing the cutouts of galaxy images from DESI Legacy Surveys online.\\

The DESI Legacy Imaging Surveys consist of three individual and complementary projects: the Dark Energy Camera Legacy Survey (DECaLS), the Beijing-Arizona Sky Survey (BASS), and the Mayall z-band Legacy Survey (MzLS). DECaLS, BASS and MzLS together include data obtained, respectively, at the Blanco telescope, Cerro Tololo Inter-American Observatory, NSF’s NOIRLab; the Bok telescope, Steward Observatory, University of Arizona; and the Mayall telescope, Kitt Peak National Observatory, NOIRLab. NOIRLab is operated by the Association of Universities for Research in Astronomy (AURA) under a cooperative agreement with the National Science Foundation. Pipeline processing and analyses of the data were supported by NOIRLab and the Lawrence Berkeley National Laboratory (LBNL). Legacy Surveys also uses data products from the Near-Earth Object Wide-field Infrared Survey Explorer (NEOWISE), a project of the Jet Propulsion Laboratory/California Institute of Technology, funded by the National Aeronautics and Space Administration. Legacy Surveys was supported by: the Director, Office of Science, Office of High Energy Physics of the U.S. Department of Energy; the National Energy Research Scientific Computing Center, a DOE Office of Science User Facility; the U.S. National Science Foundation, Division of Astronomical Sciences; the National Astronomical Observatories of China, the Chinese Academy of Sciences and the Chinese National Natural Science Foundation. LBNL is managed by the Regents of the University of California under contract to the U.S. Department of Energy. The complete acknowledgments can be found here\footnote{\url{ https://www.legacysurvey.org/acknowledgment/}}.

\bibliographystyle{scpma} 
\bibliography{main}


\end{document}